\documentclass[12pt]{iopart}

\usepackage{graphicx}
\newcommand{\bef}{\begin{figure}}
\newcommand{\eef}{\end{figure}}

\newcommand{\be}{\begin{equation}}
\newcommand{\ee}{\end{equation}}
\newcommand{\bea}{\begin{eqnarray}}
\newcommand{\eea}{\end{eqnarray}}

\begin{document}

\title[$K^{*0}$ resonance yield in baryon-rich QCD matter]{The study of hadronic rescattering on $K^{*0}$ resonance yield in baryon-rich QCD matter}

\author{Aswini Kumar Sahoo}
\address{Department of Physical Sciences, Indian Institute of Science Education and Research, Berhampur-760010, India}
\ead{aswinis19@iiserbpr.ac.in}

\author{ Subhash Singha}
\address{Institute of Modern Physics, Chinese Academy of Sciences, Lanzhou, 73000, China}
\ead{subhash@impcas.ac.cn}

\author{Md Nasim \footnote{Corresponding author}}
\address{Department of Physical Sciences, Indian Institute of Science Education and Research, Berhampur-760010, India}


\ead{nasim@iiserbpr.ac.in}
\vspace{10pt}

\begin{abstract}
The effect of hadronic rescattering on $K^{*0}$ resonance yield can be studied by measuring $K^{*0}/K$ ratio as a function of centrality or multiplicity.
This study investigates how the size of the system and the chemical composition (meson-meson versus meson-baryon interaction) of the matter formed in heavy-ion collisions impact the process of hadronic rescattering.
It is shown that existing calculation of $K^{*0}/K$ ratio, which considers the interaction of $K^{*0}$ and $K$ mesons with only light mesons in the hadronic medium (neglecting interactions with baryons), fails to explain the  measured $K^{*0}/K$ ratio at RHIC BES energies. To understand the multiplicity dependence of the $K^{*0}/K$ ratio at RHIC BES and SPS energies ($\sqrt{s_{NN}}$ $<$ 20 GeV), the Ultra Relativistic Quantum Molecular Dynamics (UrQMD) model is employed. 
UrQMD calculations suggest that for a given multiplicity, $K^{*0}/K$ is more suppressed  in Au+Au collision at $\sqrt{s_{NN}}$  =7.7 GeV, compared to that in $\sqrt{s_{NN}}$  =200 GeV.
This is possibly because of formation different type of QCD medium at 200 GeV and 7.7 GeV, and change in interaction cross-section between hadrons with change in particle type and energy.
\end{abstract}

%
%
%
%
%

 \section{Introduction}
 Ultra-relativistic nucleus-nucleus collisions generate extremely high energy densities across a large volume, providing a unique opportunity to explore and investigate the predictions of Quantum Chromodynamics (QCD) in a laboratory setting~\cite{star_white}. Several experimental observations have indicated the possible creation of a strongly coupled medium known as the quark-gluon plasma (QGP) at such high temperatures and densities~\cite{Bjorken}. 
Short-lived resonances serve as an excellent tool to study the properties of the QGP medium formed in heavy-ion collisions~\cite{Brown_resonance,Markert_resonance,Schaffner,Rapp,star_resonance}.
As the QGP medium evolves, it reaches a critical temperature known as the transition temperature. At this point, the medium undergoes a transition from a partonic phase to a hadronic phase. As the medium continues to cool, the inelastic collisions between particles come to a halt, this stage is referred to as chemical freeze-out (CFO). Following chemical freeze-out, the elastic collisions between particles cease, which is known as kinetic freeze-out (KFO)~\cite{freeze_out_1,freeze_out_2,freeze_out_3}. Resonances like $K^{*0}$ are of particular interest as their lifetime is smaller than  the duration of the medium created in heavy-ion collisions~\cite{system_life}, and they can decay, rescatter, and regenerate in between CFO and KFO, or the so-called hadronic phase.\\
The $K^{*0}$ resonance undergoes hadronic decays, specifically $K^{*0}(\overline{K^{*0}})\rightarrow K^{\pm}\pi^{\mp}$, with a branching ratio of 2/3~\cite{pdg}. Following the decay, the daughter kaons and pions have the potential to undergo rescattering interactions with other hadrons in the medium, making it challenging to reconstruct the parent resonance from its decay products. At the same time, the abundant population of kaons and pions within the medium can lead to the creation of the $K^{*0}$ resonance through pseudo-elastic scattering. This phenomenon is referred to as regeneration~\cite{reco_issue_1,reco_issue_2,reco_issue_3,reco_issue_4}.
The interplay between re-scattering and regeneration can be observed through the $K^{*0}/K$ ratio. The behavior of this ratio, specifically its suppression or enhancement, with increasing medium multiplicity $dN_{ch}/d\eta$, provides insights into the relative significance of hadronic rescattering compared to regeneration processes. A suppression of the $K^{*0}/K$ ratio with increasing medium multiplicity would suggest that hadronic rescattering dominates over regeneration, whereas an enhancement would indicate the prevalence of regeneration over rescattering.
Previous experimental measurements\cite{star_kstar_2002,star_kstar_2005,star_kstar_2008,star_kstar_2011,phenix_kstar_2014,NA49_kstar_2011,NA61_kstar_2020,NA61_kstar_2021,alice_kstar_2012,alice_kstar_2015,alice_kstar_2017,alice_kstar_2020_1,alice_kstar_2020_2,alice_kstar_2020_3,alice_kstar_2022,kstar_BES} suggest that the $K^{*0}/K$  ratios in central heavy-ion collisions are smaller compared to small collision systems like $p+p$. This implies that $K^{*0}$ and $K$  mesons undergo rescattering during the expansion of hadronic matter. The hadronic effects become larger as the size or multiplicity of the hadronic matter increases~\cite{reco_issue_2,reco_issue_4,pheno_kstar_2015,pheno_kstar_2020,pheno_kstar_2023,pheno_kstar_2018,pheno_kstar_2021}.

\begin{figure}
\begin{center}
 \includegraphics[scale=0.6]{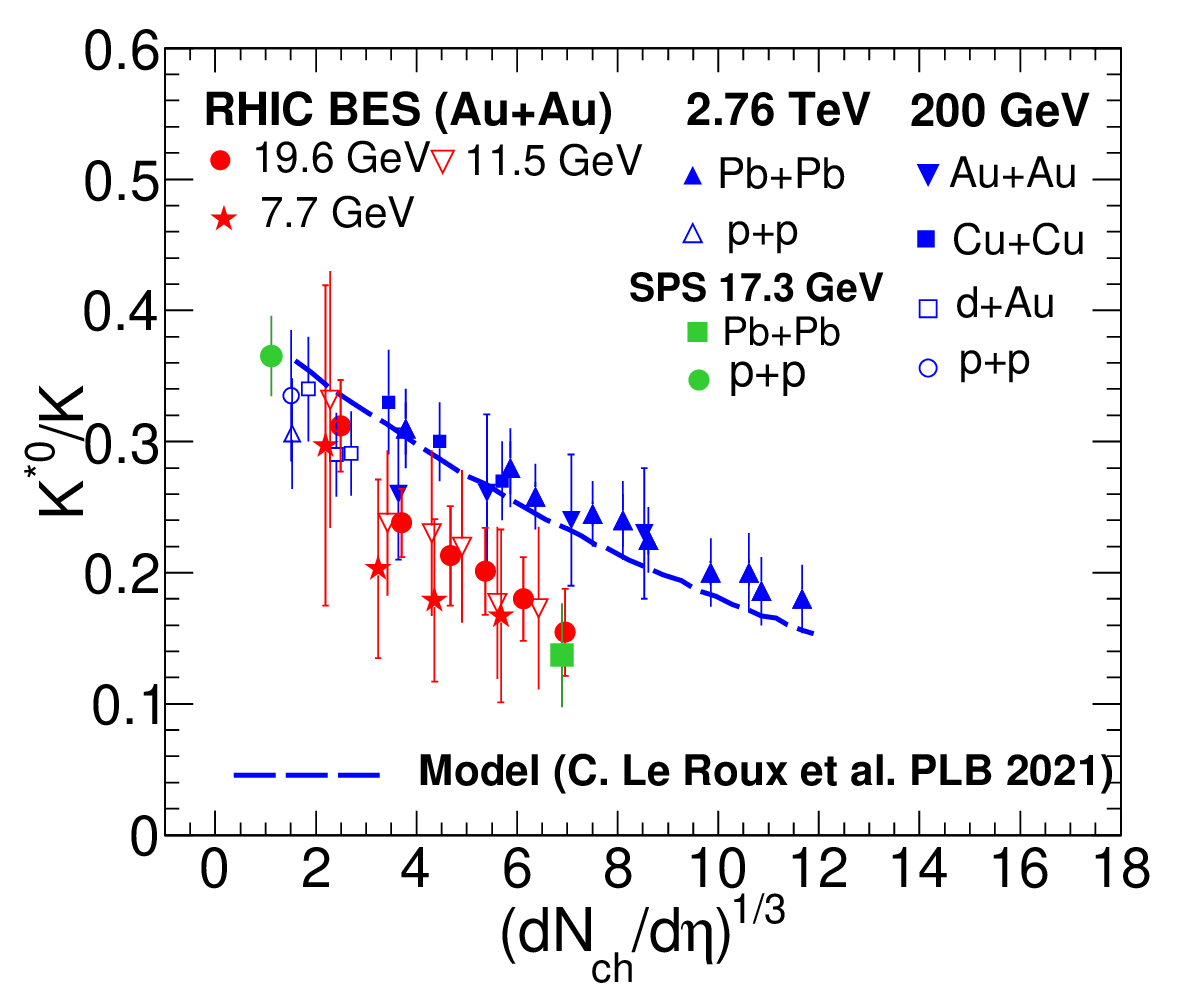}

\caption{Experimental data on $K^{*0}/K$ ratio as a function of $(dN_{ch}/d\eta)^{1/3}$ at $\sqrt{s_{NN}}$ = 7.7, 11.5 17.3, 19.6, 200 GeV and 2.76 TeV~\cite{star_kstar_2002,star_kstar_2005,star_kstar_2008,star_kstar_2011,alice_kstar_2012,alice_kstar_2015,alice_kstar_2017,alice_kstar_2020_1,alice_kstar_2020_2,alice_kstar_2020_3,alice_kstar_2022,kstar_BES}. The statistical and systematic uncertainty are added in quadrature. Values of $(dN_{ch}/d\eta)^{1/3}$ for RHIC BES energies are taken from Ref.~\cite{sqm_dndeta} and for p+p collision at 17.3 GeV, we use parametrisation given in~\cite{pp_dndeta}.
Measurement of  $K^{*0}/K$ ratio in Pb+Pb collisions at 5.02 TeV and in Au+Au collisions at 14.5, 27, 39, 62.4 GeV are also available, but  not shown for the clarity of the figure, however the conclusion remains unchanged.}
\label{intro_fig1}
\end{center}
\end{figure}

Figure~\ref{intro_fig1}  illustrates the behavior of the $K^{*0}/K$ ratio as a function of $(dN_{ch}/d\eta)^{1/3}$, which serves as a proxy for the system size, at various center-of-mass energies ranging from 7.7 GeV to 2.76 TeV. Notably, the $K^{*0}/K$ ratio exhibits a decreasing trend with increasing system size, demonstrating a smooth dependence on $(dN_{ch}/d\eta)^{1/3}$.  Figure~\ref{intro_fig1} also shows  $K^{*0}/K$ ratio from a model calculation~\cite{pheno_kstar_2021}, which considers the interaction of $K^{*0}$ and $K$ mesons with only light mesons in the hadronic medium. 
The agreement between data and model calculation~\cite{pheno_kstar_2021} at top RHIC and LHC energies indicate that rescattering is mainly dominated by meson-meson interaction at 200 GeV and above.  In this scenario, $K^{*0}$  and its decay daughters ($\pi$ and $K$)  primarily experience re-scattering with mesons resulting in meson-meson collisions dominating the rescattering process. However, at lower collision energies within the RHIC Beam Energy Scan (BES) energy regime, the meson-baryon interactions could surpass meson-meson interactions in terms of significance due to the formation of more baryon-rich matter at mid-rapidity. Furthermore, the  interaction cross-section between meson and baryon increases significantly at lower energies~\cite{KN_sigma,urqmd1}. Consequently, these meson-meson interactions at high energies and meson-baryon interactions at lower energies may lead to distinct rescattering effects on the observed $K^{*0}$ resonance. The $K^{*0}/K$ ratios in $p+p$ and peripheral A+A collisions are comparable at SPS, RHIC and LHC. Although uncertainties on the measurement at low energies (especially at 7.7 GeV) are large, $K^{*0}/K$ ratio in central collisions is consistently lower than the universal trend given by top RHIC and LHC energies.  It is also important to note that theory calculation which include only meson-meson interaction can not explain the measurement at  $\sqrt{s_{NN}}$ $<$ 20 GeV. This may suggest that meson-baryon interaction may play a dominant role at  $\sqrt{s_{NN}}$ $<$ 20 GeV.

In this study, we present a comprehensive investigation of the $K^{*0}/K$ ratio across different collision systems and energies available at RHIC, utilising the UrQMD model~\cite{urqmd1, urqmd2}. UrQMD includes both meson-meson and meson-baryon interactions. Section 2 provides a concise overview of the physics underlying the UrQMD model, along with the methodology employed for reconstructing the $K^{*0}$ resonance. The model results are presented and analysed in Section 3. Finally, a summary of the findings are provided in Section 4.


\bef
\begin{center}
\includegraphics[scale=0.8]{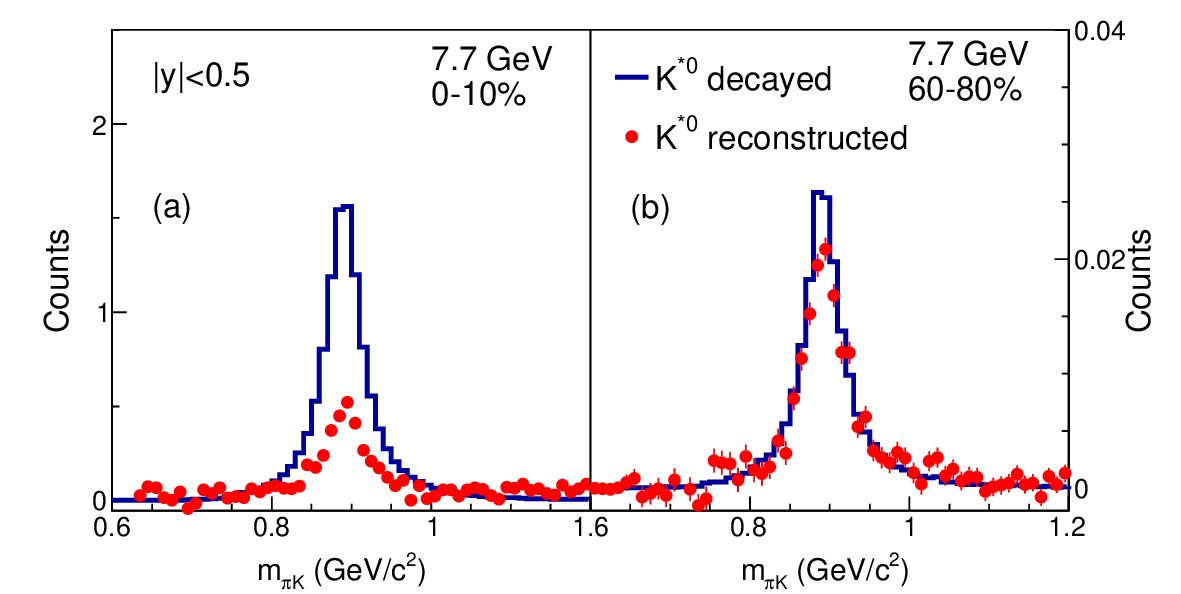}
\caption{ Invariant mass distribution of  $K^{*0}$  and its decay daughters for (a) 0-10$\%$ and (b) 60-80$\%$ Au+Au collisions at 7.7 GeV. The blue line denotes the total number of $K^{*0}$ particles that have actually decayed and the red dots denote those which could be reconstructed.}
\label{signal}
\end{center}
\eef

\section{Model description and the resonance reconstruction}
The UrQMD (Ultra-relativistic Quantum Molecular Dynamics) model~\cite{urqmd1, urqmd2},  is developed based on a microscopic transport theory, encompasses various aspects of particle production mechanisms, including resonance decays, string excitation, and fragmentation. The model focuses on the phase space description of reactions, considering the stochastic collisions of hadrons.
In the UrQMD model, the projectile and target nuclei are initialised according to a Woods-Saxon profile in coordinate space. The stochastic collisions between hadrons are performed similarly to the original cascade models, until the collision criteria is satisfied. When the relative distance $d_{trans}$ between two particles in three-dimensional configuration space becomes smaller than a critical distance, a collision occurs. The critical distance $d_0$ is determined by the corresponding total cross section $\sigma_{tot}$, which depends on the center-of-mass energy $\sqrt{s}$, the species and quantum numbers of the incoming particles.
The UrQMD model incorporates more than 50 baryon species, including nucleons, delta and hyperon resonances with masses up to 2.25 GeV/$c^2$, and 45 meson species, including strange meson resonances. The model also includes the corresponding antiparticles and isospin-projected states. The interactions in the model encompass baryon-baryon, meson-baryon, and meson-meson interactions, allowing for a comprehensive description of the dynamics and interactions within the hadronic system~\cite{urqmd1, urqmd2}.\\
\begin{figure*}[ht]
  \includegraphics[scale=0.8]{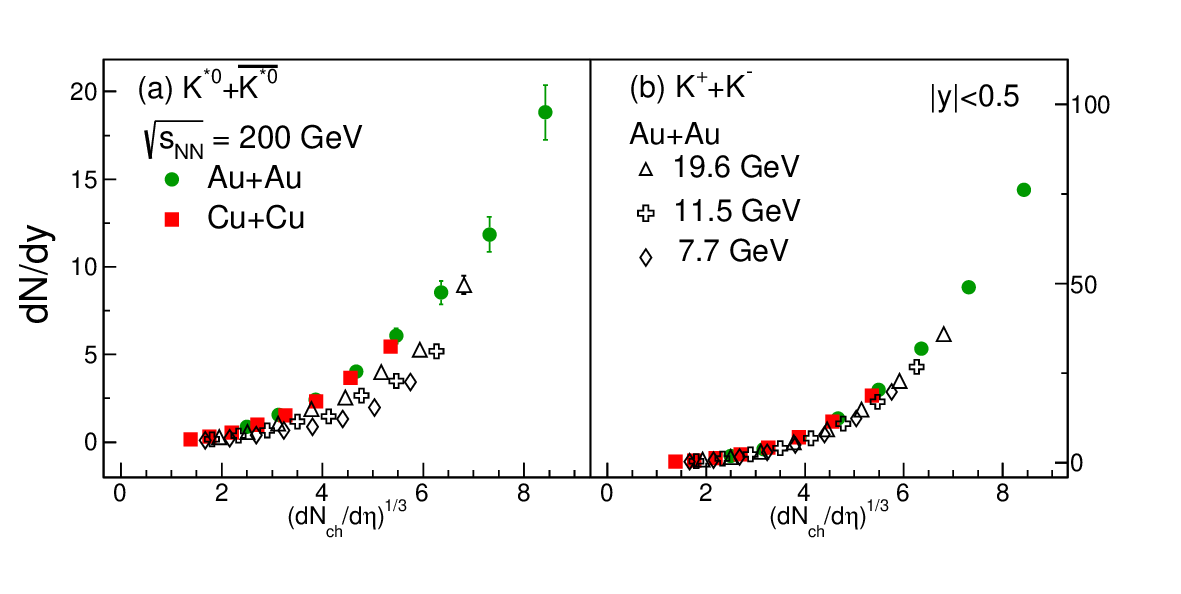}
\caption{The $p_{T}$ integrated yield of $K^{*0}$ and charged kaons as a function of $(dN_{ch}/d\eta)^{1/3}$ for various collision systems and beam energies from  the UrQMD calculation. 
The symbols indicate different collision results: the green circle and red square represent Au+Au and Cu+Cu collisions at $\sqrt{s_{NN}}$ = 200 GeV, respectively. The open triangle, open plus, and open diamond denote results from Au+Au collisions at $\sqrt{s_{NN}}$ = 7.7, 11.5, and 19.6 GeV, respectively.
}
\label{yield}
\end{figure*}


In our investigation, we employed version 2.3 of the UrQMD model, which is publicly available. We ran the model using a cascade approach. In this study, approximately 4 million events for each configuration were generated for minimum-bias A+A collisions. \\
In this model, the masses of resonances are distributed according to the Breit-Wigner function, which allows for the simulation of their decay processes.
During the simulation, the decay of unstable particles was enabled to account for their natural decay processes. 
For the reconstruction of short-lived particles such as $K^{*0}$, we implemented a method similar to that used in experimental data analysis. This involved accumulating invariant mass distributions for different combinations of daughter particles and applying proper treatments to eliminate background contributions through track rotation technique. 

Figure~\ref{signal} illustrates the invariant mass distribution of  $K^{*0}$  and its decay daughters. The blue line represents the invariant mass distribution of $K^{*0}$ particles that have actually decayed following a Breit-Wigner distribution. The red data points indicate the reconstructed invariant mass distribution of $K^{\pm}\pi^{\mp}$ pair combinations.
The number of reconstructed particles is less than those that actually decayed, indicating loss of  $K^{*0}$ signal due to  in-medium interactions of its daughters particles with in-medium particles. The effect of in-medium interactions or hadronic re-scattering is more pronounced in 0-10$\%$ central collisions compared to  60-80$\%$ peripheral collisions. The yield of the $K^{*0}$ resonance was then estimated by integrating the resonance mass peak and correcting for the corresponding branching ratio.
The $K^{*0}$ and $\overline{K^{*0}}$ is combined and denoted as $K^{*0}$, unless specified. Similarly the charged kaons are combined and denoted as K.

\bef
\includegraphics[scale=0.41]{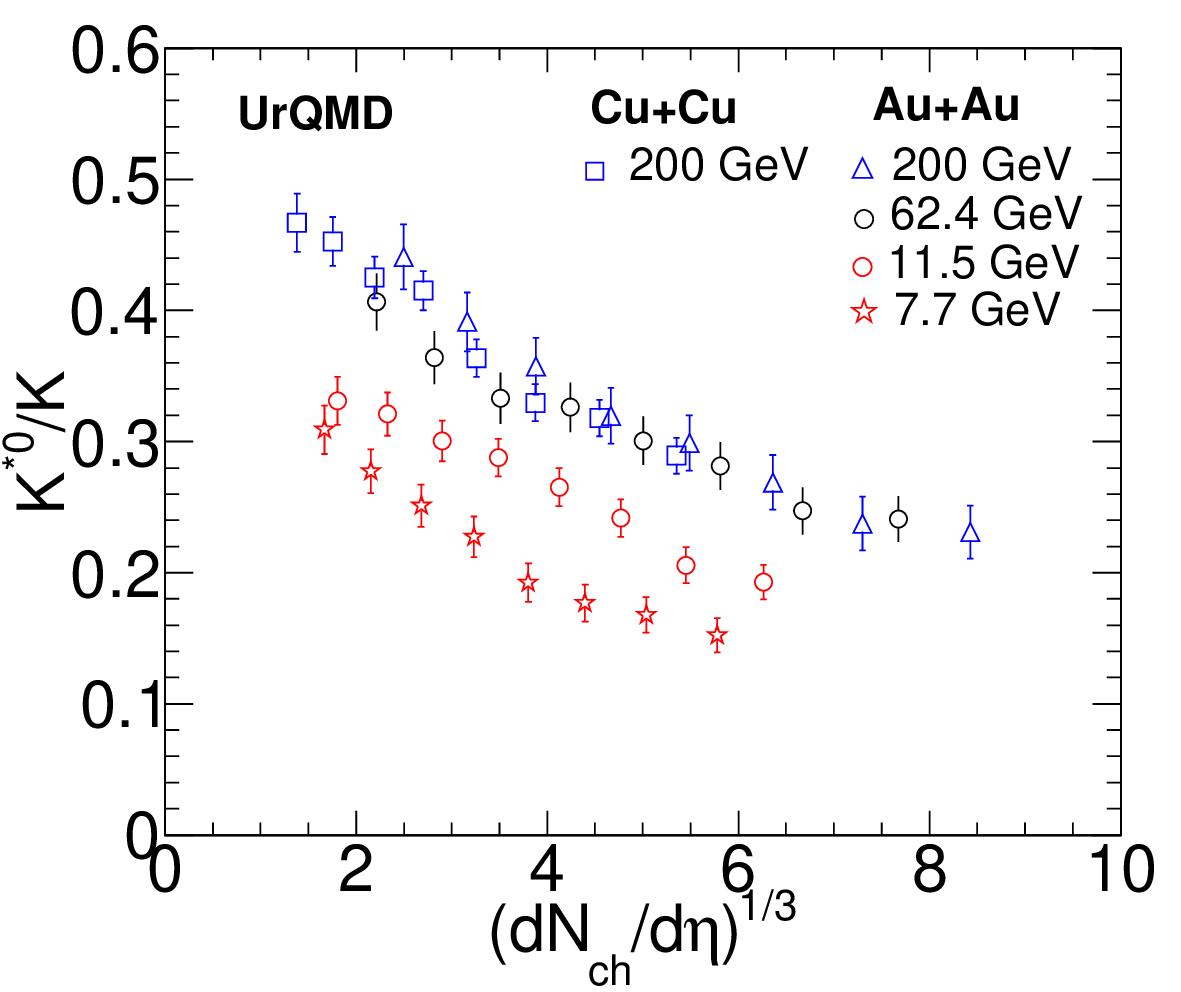}
\includegraphics[scale=0.39]{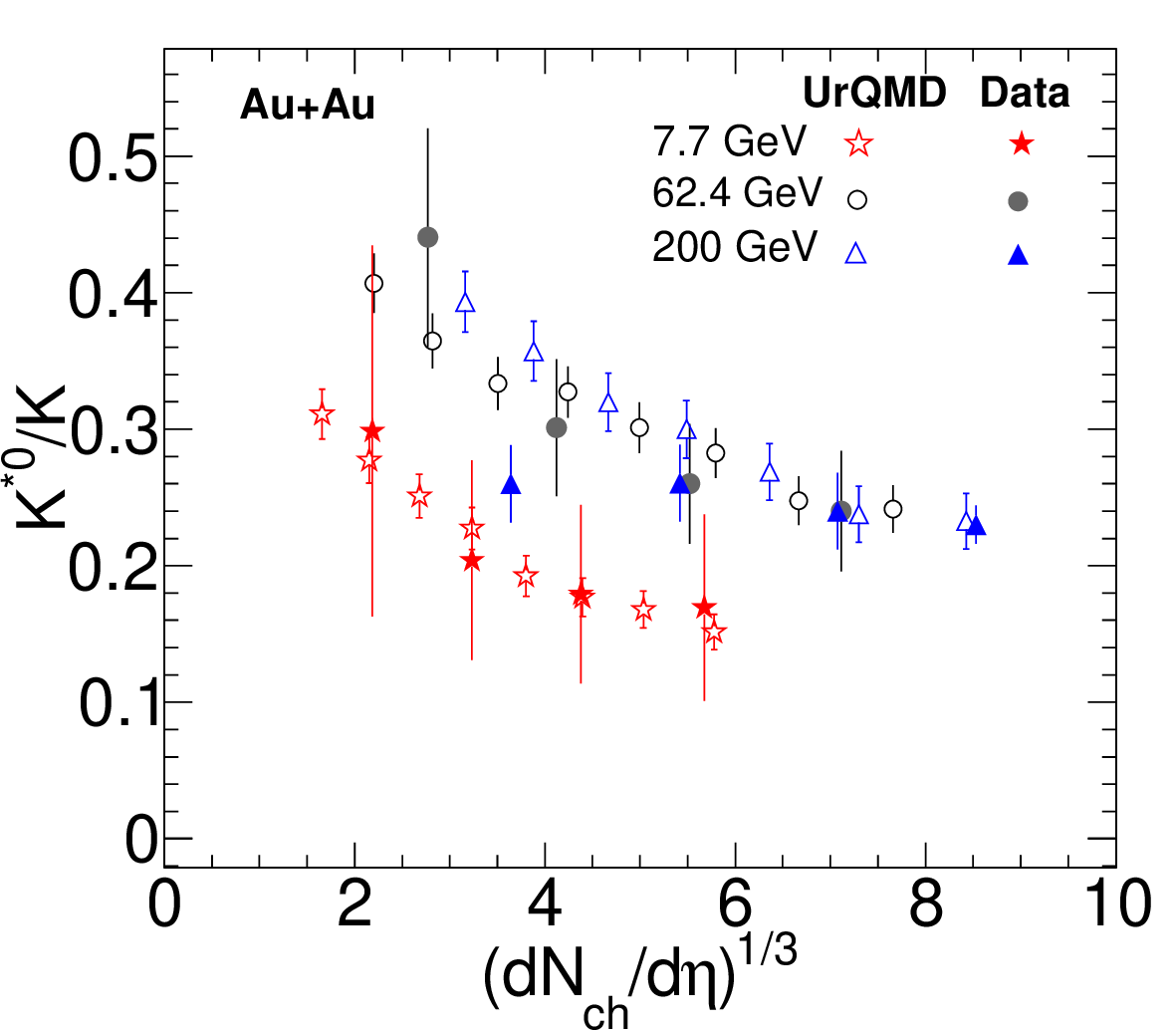}
\caption{Left: The $K^{*0}/K$ ratio calculated at mid-rapidity as a function of $(dN_{ch}/d\eta)^{1/3}$ for various collision systems and beam energies from  the UrQMD calculation. Right: Comparison of $K^{*0}/{K}$ ratios from UrQMD with STAR data as a function of $(dN_{ch}/d\eta)^{1/3}$ in Au+Au collisions at 7.7 GeV, 62.4 and 200 GeV.}
\label{ratio}
\eef

\section{Results and Discussions}

Figure~\ref{yield} illustrates the integrated yield (dN/dy) of $K^{*0}$ mesons and charged kaons as a function of $(dN_{ch}/d\eta)^{1/3}$. The model calculations are performed within the rapidity range $|y| < 0.5$ for two different collision systems: Au+Au and Cu+Cu, at a center-of-mass energy of $\sqrt{s_{NN}}$ = 200 GeV. Additionally, calculations are shown for Au+Au collisions at different  center-of-mass energies: $\sqrt{s_{NN}}$ = 7.7, 11.5, and 19.6 GeV.

The results indicate that the yield of charged kaons increases with multiplicity across all collision systems and beam energies. Furthermore, for a given multiplicity bin, the yield is nearly independent of both the collision system and beam energies.
In contrast, the yield of $K^{*0}$ mesons shows a different trend when plotted as a function of multiplicity. The yield of $K^{*0}$ mesons varies with different beam energies, particularly at lower beam energies, for a given multiplicity bin. These findings at lower beam energies suggest that additional factors, such as loss due to re-scattering, influence the measured yield.

Figure ~\ref{ratio} (left panel) presents the system size dependence of the $K^{*0}/K$ ratio as a function of $(dN_{ch}/d\eta)^{1/3}$ in the UrQMD model. 
At a center-of-mass energy of $\sqrt{s_{NN}}$ = 200 GeV and 62.4 GeV, the ratios exhibit an approximate multiplicity scaling irrespective of the collision species as observed in data measured at top RHIC energies. However, as the collision energy decreases, the $K^{*0}/K$ ratio is significantly lower compared to that at 200 and 62.4 GeV. This deviation suggests that even at the same multiplicity or system size, the effect of re-scattering is not the same at $\sqrt{s_{NN}}$ = 200 GeV compared to lower energies such as 7.7 GeV. Alternatively, this difference could be due to changes in the chemical composition of the hadronic medium at mid-rapidity between RHIC BES and the top RHIC energy. \\

Figure ~\ref{ratio} (right panel)  show the comparison of $K^{*0}/{K}$ ratios derived from UrQMD with the experimental data obtained by the STAR experiment. The red markers represent the experimental data, while the blue open markers depict the UrQMD calculations. UrQMD successfully accounts for the data at 7.7 and 62.4 GeV, and only for central collisions at 200 GeV. 

 
\bef
\begin{center}
\includegraphics[scale=0.8]{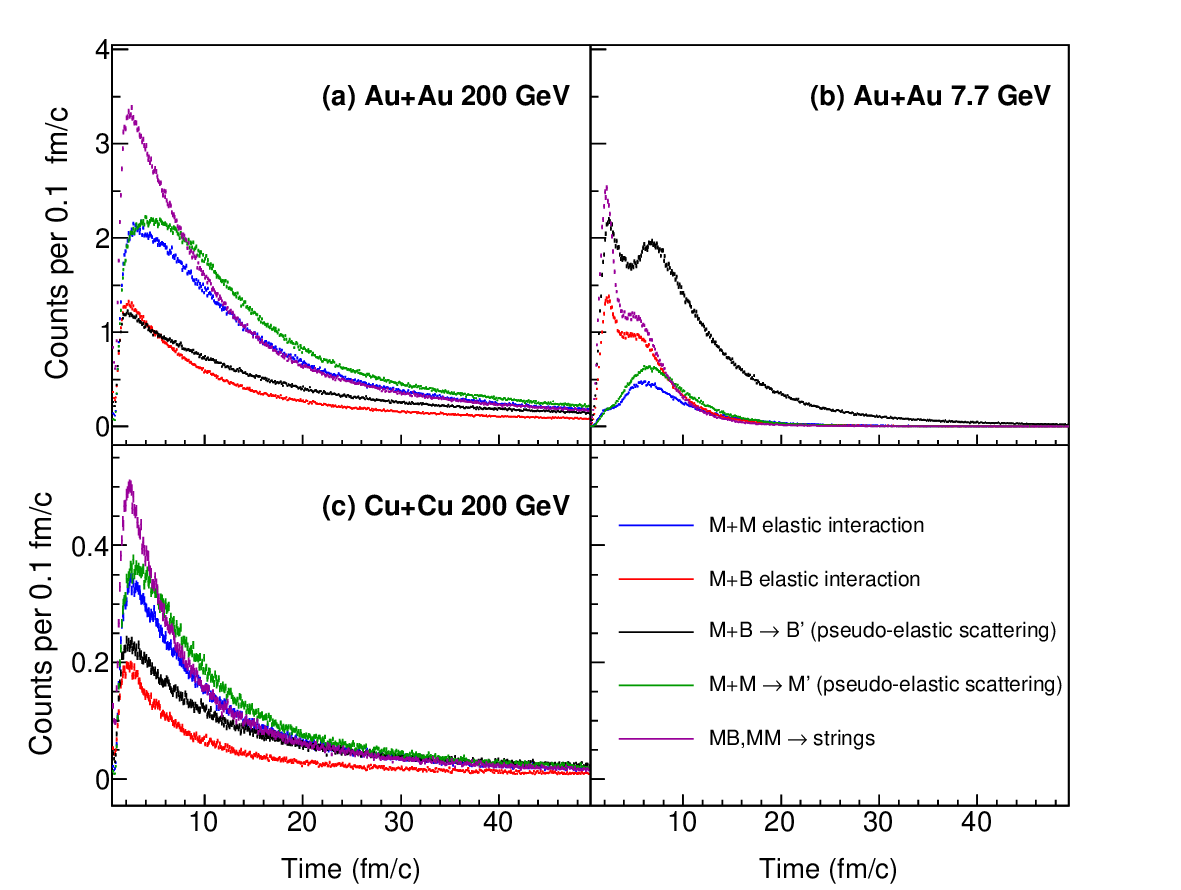}
\caption{The time-differential distribution of hadronic interaction for meson-meson(M+M) and meson-baryon(M+B) pairs for $\sqrt{s_{NN}}$= 7.7 and 200 GeV (Au+Au and Cu+Cu) from UrQMD model. The elastic scatterings between meson-meson and meson-baryon pairs are represented by blue and red lines, respectively. Green and black lines indicate pseudo-elastic scatterings between MM and MB pairs, respectively. Inelastic interactions involving meson-meson or meson-baryon pairs (MM, MB to strings) are depicted in magenta. }
\label{interaction}
\end{center}
\eef


\bef
\begin{center}
\includegraphics[scale=0.8]{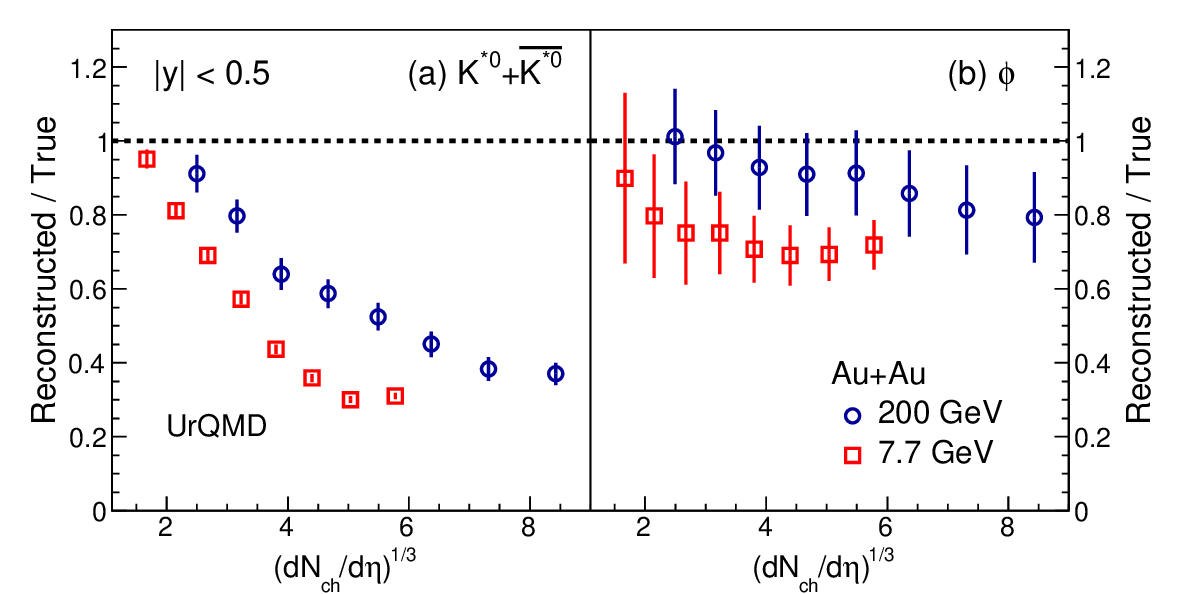}
\caption{The ratio between number of reconstructed  and number of true resonances  as a function of $(dN_{ch}/d\eta)^{1/3}$ from UrQMD model in Au+Au collisions at $\sqrt{s_{NN}}$= 7.7 and 200 GeV.
The left  and right panels are for $K^{*0}$  and $\phi$ resonances, respectively.}
\label{double_ratio}
\end{center}
\eef

\newpage

The UrQMD model serves as a comprehensive framework for examining the complete evolution of hadronic interactions concerning propagation time or medium evolution time($\tau$). 

Figure~\ref{interaction} illustrates the interaction probabilities between meson-meson (M+M) and meson-baryon (M+B) pairs as a function of medium evolution time.
The presented distributions pertain to two collision energies, specifically $\sqrt{s_{NN}}$ = 7.7 GeV (Au+Au) and 200 GeV(Au+Au and Cu+Cu). In this case, only the distributions of meson-meson and meson-baryon interactions are shown since the $K^{*0}$ resonance decays into $\pi$ and $K$ particles.
 Inelastic interactions involving meson-meson or meson-baryon pairs (MM, MB to strings) are depicted in magenta. In UrQMD output, inelastic scatterings between meson-meson and meson-baryon pairs cannot be individually distinguished.  
The elastic scatterings between meson-meson and meson-baryon pairs are represented by blue and red lines, respectively. Additionally, pseudo-elastic scatterings, which result in the formation of an intermediate resonance state, are also presented in Fig~\ref{interaction}. Green and black lines indicate pseudo-elastic scatterings between MM and MB pairs, respectively.
At  top RHIC energies the medium produced at mid-rapidity is predominantly composed of pions, making pion-pion interactions or meson-meson elastic and pseudo-elastic scatterings the dominant processes that can influence the decay products of K* resonances. Figure~\ref{interaction} shows,  at  $\sqrt{s_{NN}}$ = 200 GeV,  for elastic and pseudo-elastic scattering, meson-meson interactions dominates over  baryon-meson interactions in both Au+Au and Cu+Cu  and we observed that $K^{*0}/K$ ratio shows smooth multiplicity scaling in all such collisions system.
Given the large pion-nucleon cross section at low center-of-mass energies, meson-baryon interactions are particularly significant. At these energies, meson-baryon interactions are largely governed by  pseudo-elastic scatterings, leading to the formation of intermediate resonances. As shown in Ref~\cite{{pdg}} and implemented in UrQMD, the measured pseudo-elastic cross-section for the formation of the $\Delta^{++}$ resonance state in the $\pi^{+}+p$ system is about 200 mb, much higher than the approximately 120 mb pseudo-elastic cross-section for the formation of the $\rho$ meson in the $\pi+\pi$ system.
Similarly, the $K^{-}+p$ total cross-section can reach up to 120 mb at low kaon beam momentum, compared to only 20 mb at higher beam momenta~\cite{urqmd1,urqmd2,smash}.
At $\sqrt{s_{NN}}$ = 7.7 GeV, the average momentum of produced particles is smaller compared to those at 200 GeV, leading to an expectation that meson-baryon elastic and pseudo-elastic interactions will be more relevant than meson-meson elastic and pseudo-elastic interactions. Figure~\ref{interaction} clearly demonstrates that at $\sqrt{s_{NN}}$ = 7.7 GeV, both MB elastic and MB pseudo-elastic interaction rates are higher than those of MM elastic and pseudo-elastic interactions.

Within the UrQMD model, it is feasible to calculate the true number of resonances that decayed during the evolution of hadronic medium. 
This offers a chance to examine the influence of hadronic re-scattering on the generation of short-lived resonance particles. 
To understand the nature of rescattering at 7.7 GeV and 200 GeV, we have computed the  ratio of reconstructed by true resonances as a function of $(dN_{ch}/d\eta)^{1/3}$.

Figure~\ref{double_ratio}(left panel) shows the ratio between the number of reconstructed  and  true $K^{*0}$ resonances as a function of $(dN_{ch}/d\eta)^{1/3}$ for Au+Au collisions at both 7.7 GeV and 200 GeV. The results indicate a more significant suppression in the $K^{*0}$ yield at 7.7 GeV compared to 200 GeV during the hadronic evolution. This observation emphasizes that, for a given multiplicity, the rescattering effect is more pronounced at 7.7 GeV than at 200 GeV, emphasizing the importance of meson-baryon interaction in determining the $K^{*0}/K$ ratio at lower energies.
Additionally, in Fig.~\ref{double_ratio}(right panel), the  ratio  $\phi(reconstructed)/\phi(true)$  is plotted as a function of $(dN_{ch}/d\eta)^{1/3}$ for Au+Au collisions at both 7.7 GeV and 200 GeV. The ratio for $\phi(reconstructed)/\phi(true)$ shows  weaker suppression compared to $K^{*0}$. This behavior can be attributed to the fact that the $\phi$ meson is minimally influenced by hadronic re-scattering effects, given its long lifetime (42 fm/c). Furthermore, 
the ratio $\phi(reconstructed)/\phi(true)$ is more suppressed at 7.7 GeV compared to 200 GeV.

\section{Summary}
In summary, we investigated the dependence of $K^{*0}$ production on system size and collision energy in various collision systems and beam energies available at RHIC, employing the UrQMD model.  At the top RHIC, the $K^{*0}/{K}$ ratio  exhibited an approximate scaling with collision multiplicity. However, as the collision energy decreased, the $K^{*0}/{K}$ ratio found to be significantly lower than that at top RHIC energy.
The agreement between UrQMD model calculation of  $K^{*0}/K$ ratio and data looks better in Au+Au collision at $\sqrt{s_{NN}}$  =7.7 GeV compared to that in 200 GeV. It was also found that, $K^{*0}(reconstructed)/K^{*0}(true)$ or  $\phi(reconstructed)/\phi(true)$ is more suppressed  in Au+Au collision at $\sqrt{s_{NN}}$  =7.7 GeV, compared to that in $\sqrt{s_{NN}}$  =200 GeV. The UrQMD model suggests that meson-baryon interactions may play a more prominent role in a more baryon-rich environment at lower energy regimes, contrasting with the dominance of meson-meson interactions at higher energies. To gain further insights into rescattering in the BES energy range, it would be valuable to compare these predicted results with measurements from high-statistics BES-II data. This comparison can provide a better understanding of the dynamics of the medium and the interplay between collision energy, system size, and the chemical composition when studying the hadronic rescattering . This study can be further extended using hybrid urqmd model which include  an intermediate hydrodynamic stage.


\section*{References}

\end{document}